\documentstyle[epsfig]{aipproc}

\def\erg{{\rm\thinspace erg}}

\def\keV{{\rm\thinspace keV}}
\def\km{{\rm\thinspace km}}

\def\s{{\rm\thinspace s}}

\def\ergps{\hbox{$\erg\s^{-1}\,$}}

\def\kmps{\hbox{$\km\s^{-1}\,$}}

\begin{document}
\title{X-rays and accretion discs as probes of the strong gravity of
black holes}

\author{A.C. Fabian}
\address{Institute of Astronomy, Madingley Road, Cambridge CB3 0HA, UK}

\maketitle

\begin{abstract}
The observations and interpretation of broad iron lines in the X-ray
spectra of Seyfert 1 galaxies are reviewed. The line profiles observed
from MCG--6-30-15 and NGC\,3516 show extended red wings to the line
explained by large gravitational redshifts. The results are consistent
with the emission expected from an X-ray irradiated flat accretion
disc orbiting very close to a black hole. Results from XMM-Newton and
Chandra are presented and the possibility of broad oxygen lines
discussed.
\end{abstract}

\section*{Introduction}
The optical (e.g. Ghez et al 2000; Eckart et al 1997; Gebhardt et al
2000; Ferrarese \& Merritt 2000) and radio (e.g. Miyoshi et al 1995)
evidence for compact central masses in the nuclei of many nearby
galaxies is now very clear. The compact objects are consistent with
being supermassive black holes. The data do not however probe the
strong gravity regime of these holes, since the observed stars and gas
orbit at more than 40,000 gravitational radii from the centre.

Much of the X-ray emission from accreting black holes should emerge
from within a few 10s of gravitational radii. Aspects of this emission
can provide us with a probe of the strong gravity of black holes. In
particular the rapid variability, or in some cases quasi-periodic
oscillations, and soft thermal emission from an accretion disc
demonstrates the small size of the X-ray emission region. 

Here I concentrate on the broad iron emission lines seen in the
spectra of some Seyfert 1 galaxies. The line profile and its
variability can reveal the geometry of the accretion flow, and strong
gravitational effects, within a few gravitational radii of the black
hole. The dominant expected line is that due to iron, as shown in
Fig.~1 (George \& Fabian 1991; Matt, Perola \& Piro 1991). The
spectrum shown is the result of a Monte-Carlo simulation of the
emergent spectrum when a flat surface has been irradiated with a
power-law X-ray spectrum. The gas is assumed to be neutral, although
the important issue here is that the metals retain their K and L-shell
electrons. The major emission line is iron K-$\alpha$, due to its
relatively high abudance (a cosmic mix is assumed) and high
fluorescent yield (which increases as $Z^4$). Doppler broadening of
this emission line, produced by the irradiation of the surface of an
accretion disc by hard X-rays from its corona, together with thr
special relativistic effects of beaming and the transverse doppler
effect and the general relativistic effect of gravitational redshift,
lead to the observed profile being very broad and skew (Fabian et al
1989: Figs.~2, 3). The 'blue horn' of the line is most sensitive to
the inclination angle of the disc (i.e. the importance of doppler
broadening) and the 'red wing' is most sensitive to the inner radius
(i.e. the gravitational redshift).

\begin{figure}[] 
\centerline
{\epsfig{file=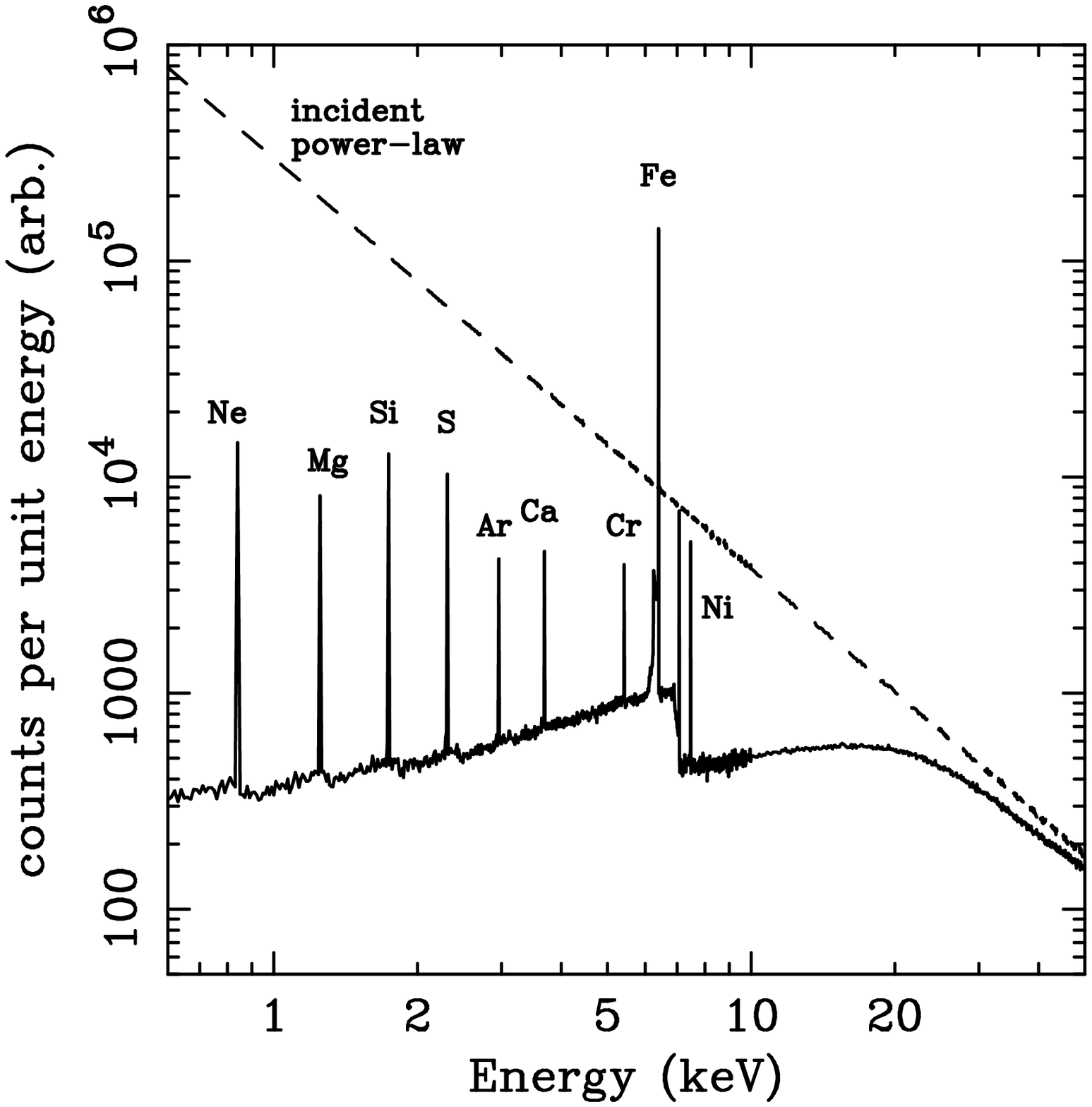,width=3in,height=5in,angle=270}}
\vspace{10pt}
\caption{Monte Carlo simulation (by C. Reynolds) of the reflected
continuum plus fluorescent line emission from a neutral slab of gas
with comsic abundances. The incident power-law continuum is indicated
by the dashed line. In practice the sum of the incident and reflected
spectra are usually seen.}
\end{figure}

\begin{figure}[] 
\centerline
{\epsfig{file=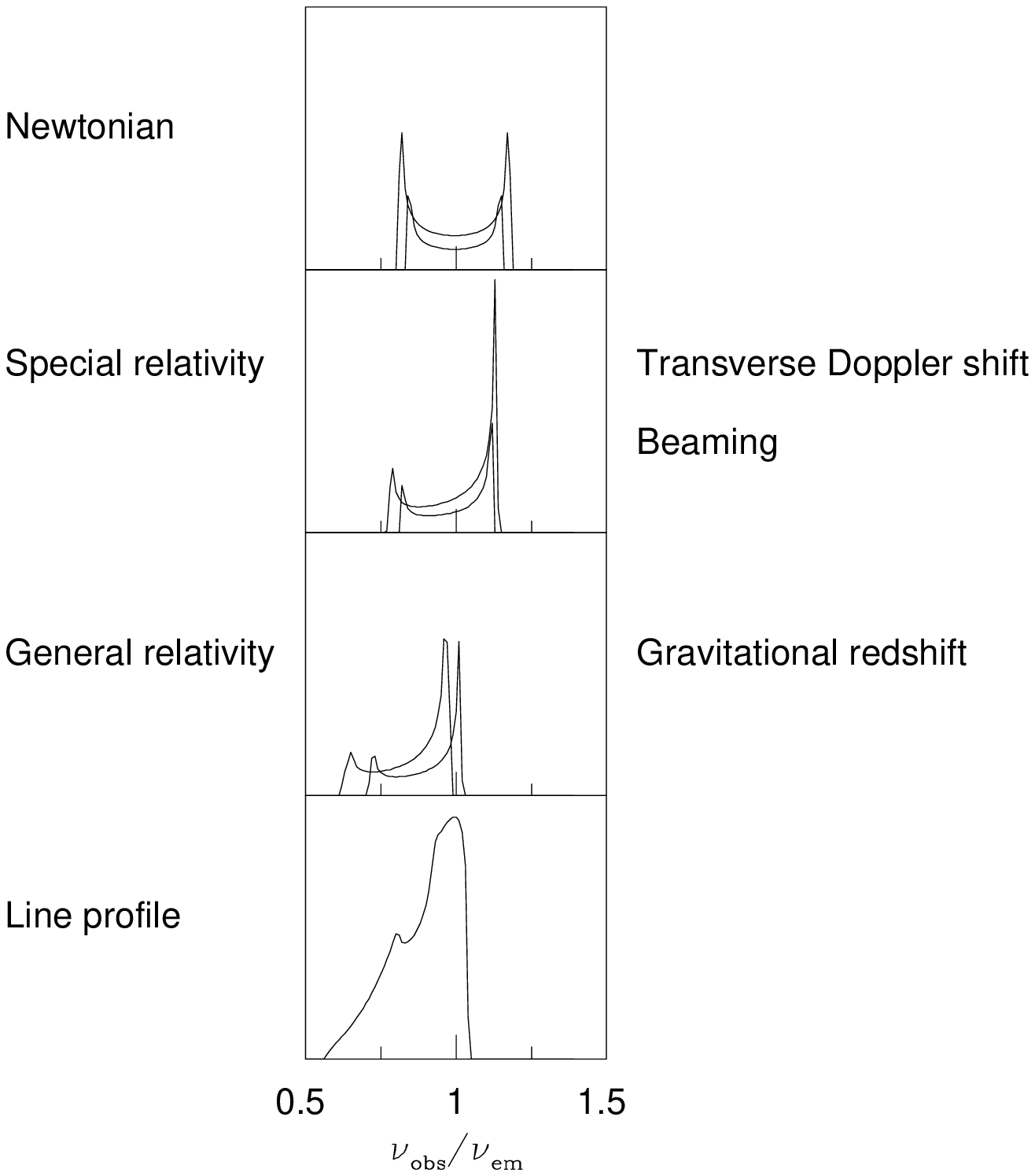,width=6in,height=3in}}
\vspace{10pt}
\caption{Line broadening from an intrinsically narrow line emitted
from two radii in an accretion disc (from Fabian et al 2000). The
lowest panel show the result obtained by summing many disc radii,
weighted by the expected emissivity.}
\end{figure}

\begin{figure}[] 
\hbox{
{\epsfig{file=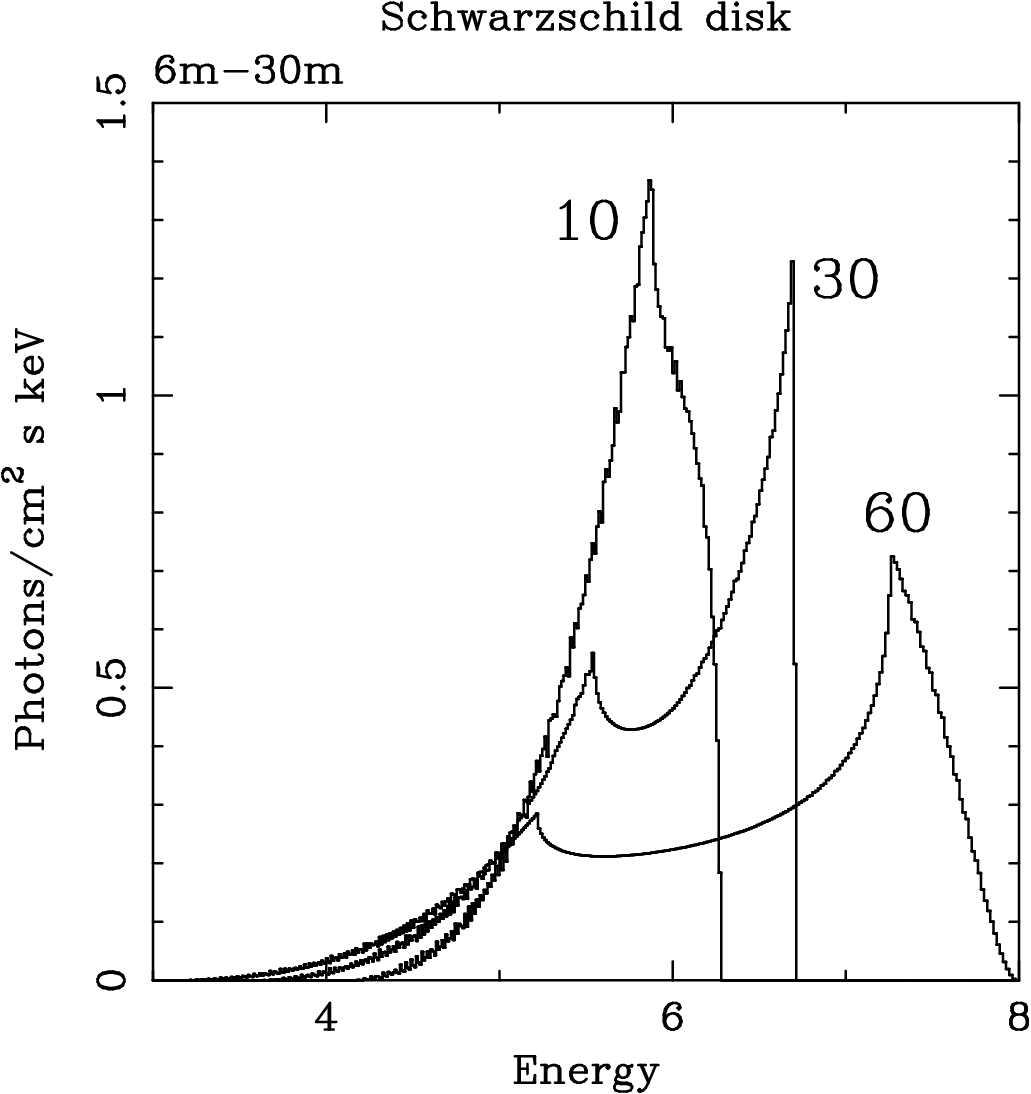,width=3in,angle=270}}
{\epsfig{file=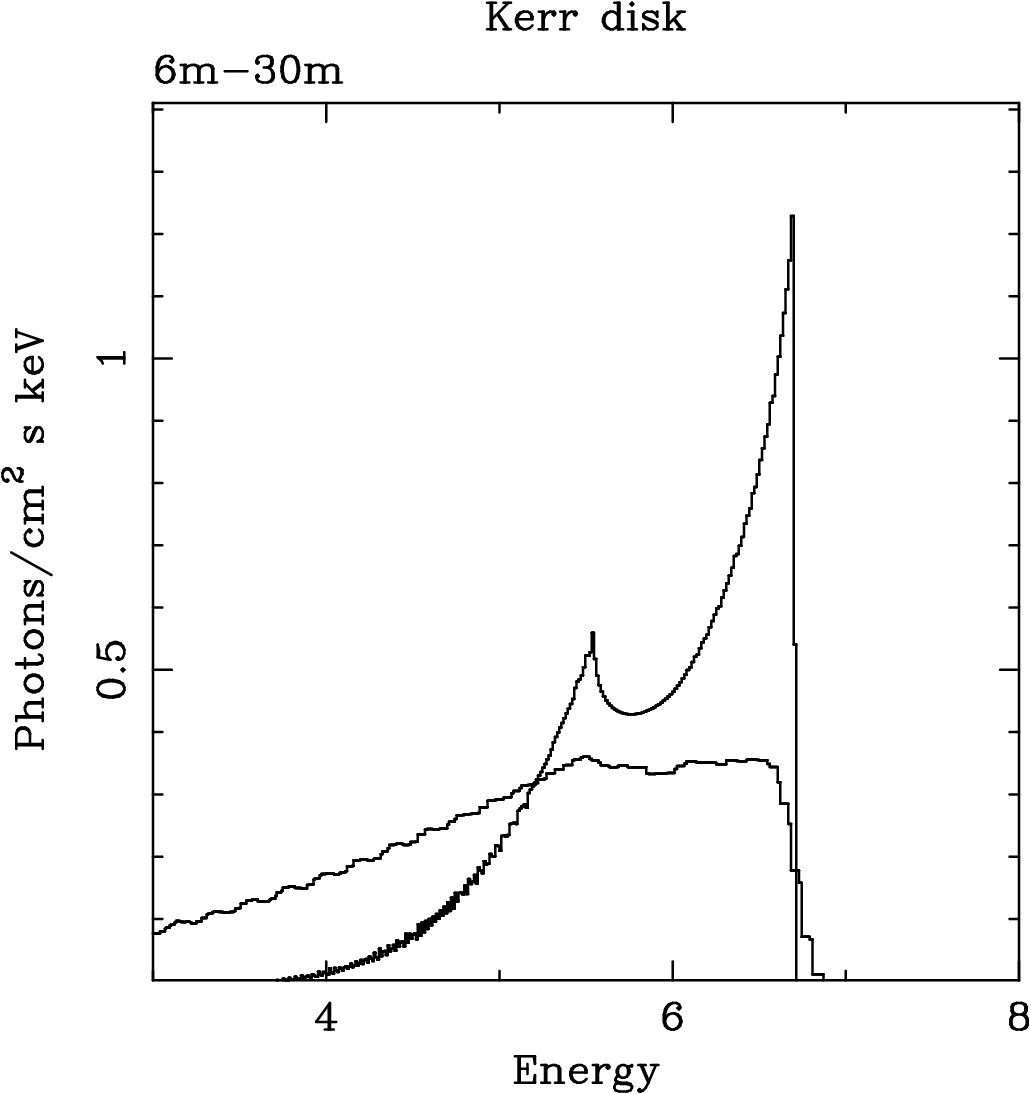,width=3in,angle=270}}}
\vspace{10pt}
\caption{Model spectra produced by the iron K$\alpha$ 6.4~keV line
emitted from an accretion disc at radii (Left) $6-30$r$_{\rm g}$
around a non-spinning (Fabian et al 1989) and (Right) $6-30$r$_{\rm
g}$ and (broader profile) $1.25-30$r$_{\rm g}$ around a
maximally-spinning (Laor 1990) Kerr black hole. }
\end{figure}

\section{Observations of broad iron lines}
A clear broad iron line was first seen with data from a 1994 ASCA
observation of the Seyfert 1 galaxy MCG--6-30-15 (Tanaka et al 1995:
Fig.~4). This was the result of a 4.5 day exposure on the source and
is shown with the dominant power-law continuum removed. The data are
well fit (Fabian et al 1995) by emission from an accretion disc
stretching from about 6 gravitational radii (i.e. $6r_{\rm
g}=6GM/c^2$) to about 40 $r_{\rm g}$ and an inclination of about 30
deg. The surface emissivity falls as radius $r^{-3}$. A similar clear
line has also been seen in NGC\,3516 (Nandra et al 1999: Fig.~4). The
shape of these lines is entirely consistent with the emitting region
being a flat disc in Keplerian rotation close to a black hole.

K. Nandra (priv. comm.) has compiled the line spectra of several more
Seyfert 1 galaxies (Fig.~5) which generally are weaker in flux, or
have shorter exposures than for MCG--6-30-15. A broad red wing is seen
in most. Further work on the strength and width of iron lines in
Seyferts is reported by Lubi\'nski
\& Zdziarski (2000).

\begin{figure}[] 
\hbox{
{\epsfig{file=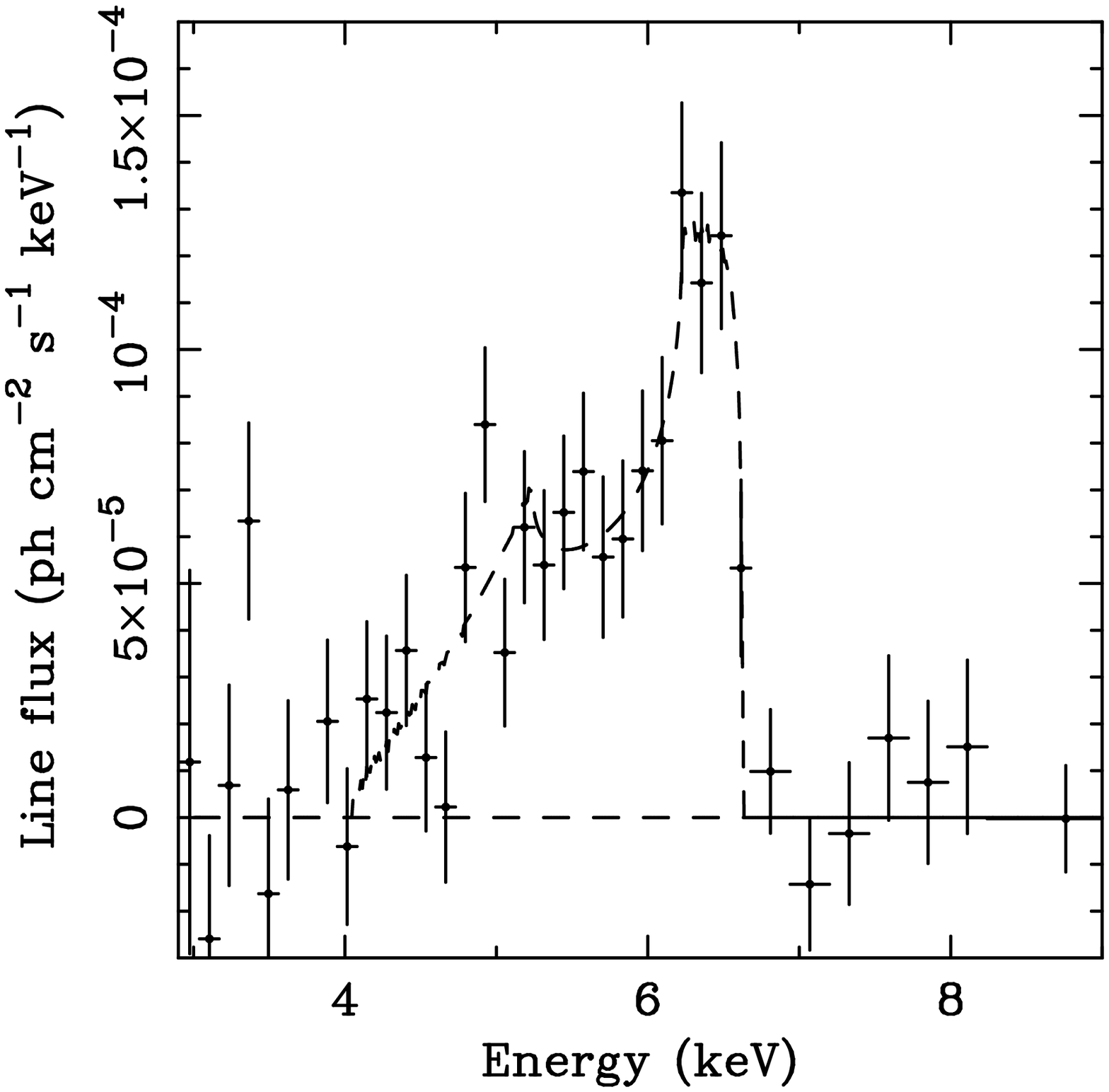,height=2.5in}}
{\epsfig{file=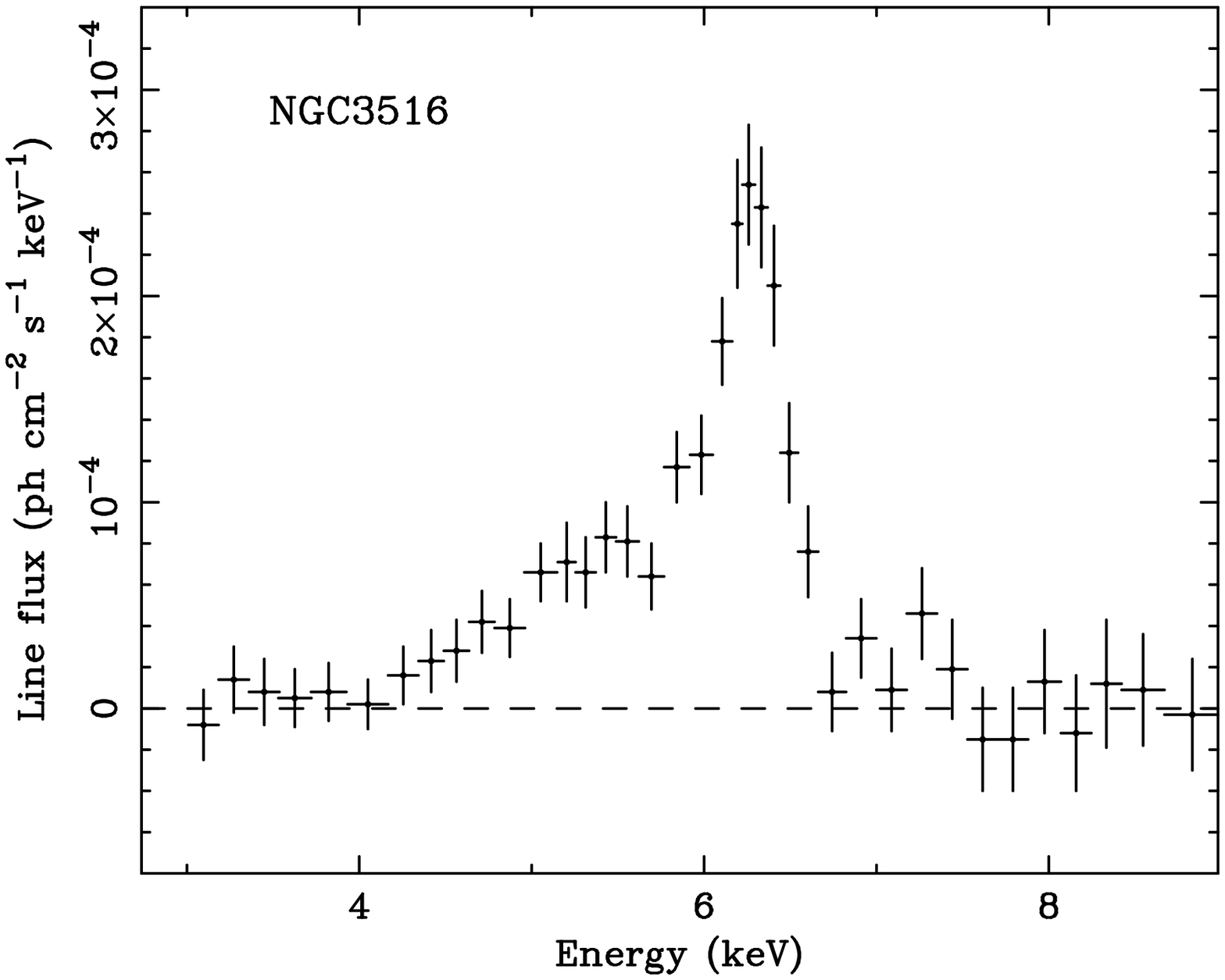,height=2.5in}}
}
\vspace{10pt}
\caption{Observed profiles of broad iron lines; (Left) 
MCG--6-30-15 from Tanaka et al (1995) and (Right) NGC\,3516 from
Nandra et al (1999).  }
\end{figure}

\begin{figure}[] 
\centerline{\epsfig{file=fig_prof_col.ps,height=2.5in,width=2.5in}}
\vspace{10pt}
\caption{Line profiles for a selection of Seyfert 1 galaxies made by
K. Nandra (prov. comm.). }
\end{figure}

\begin{figure}[] 
\hbox{{\epsfig{file=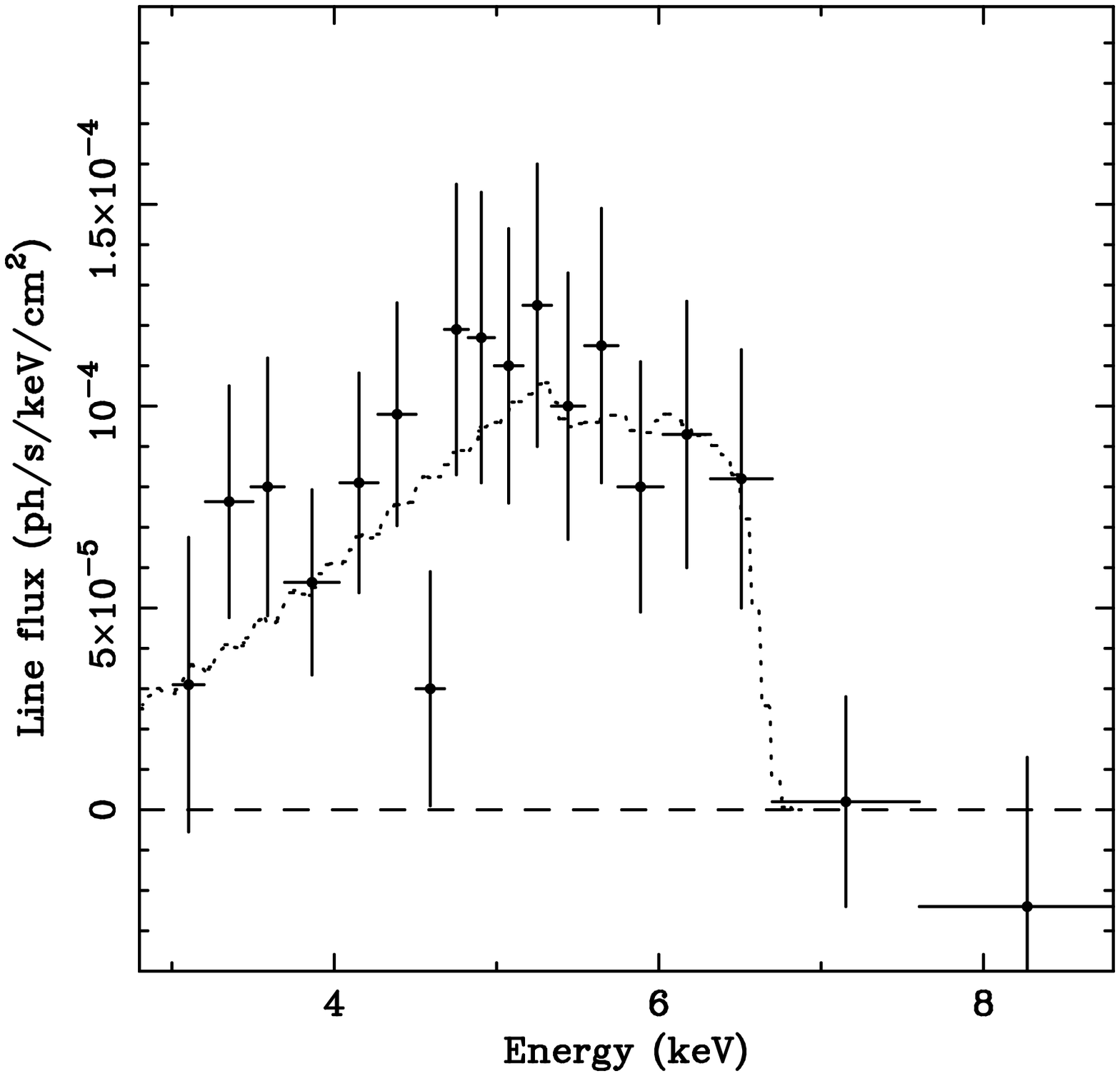,width=2.5in,angle=270}}
{\epsfig{file=iondiscb.ps,width=2.5in,angle=270}}}
\vspace{10pt}
\caption{(Left) line profile druing the deep minimum of the ASCA 1994
observation (Iwasawa et al 1996). The fitted profile is for a maximal
Kerr black hole. (Right) model profile for a
relativistically-broadened, constant density ionized disc model fitted
to the time-averaged 1994 ASCA data. Compare with Fig.~4, left panel).
}
\end{figure}

Occasionally the iron line in MCG--6-30-15 changes its profile, as in
a deep minimum (Iwasawa et al 1996) or after a flare (Iwasawa et al
1999), and appears to stretch more to the red. This is possible
evidence that the black hole is spinning rapidly (Iwasawa et al 1996;
Dabrowski et al 1997), although matter streaming from the innermost
stable orbit around a non-spinning black hole might mimic this result
(Reynolds \& Fabian 1997; Young et al 1998).

\begin{figure}[t] 
\centerline{\epsfig{file=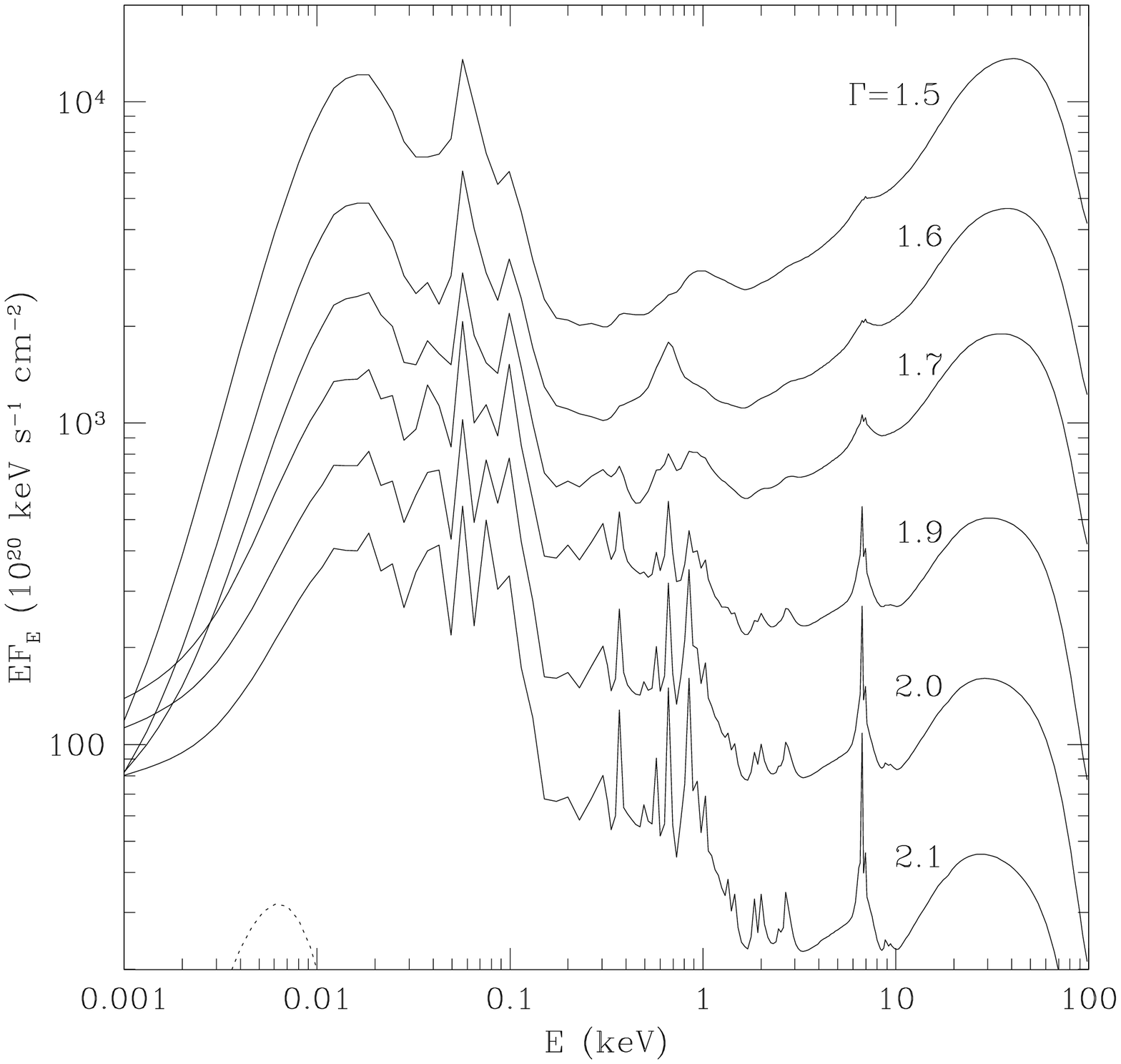,height=3in,width=4.in}}
\vspace{10pt}
\caption{Reflection spectra from ionized accretion disc models,
before relativistic blurring; see Ballantyne et al (2001) for details.
The observed spectrum will usually need the addition of the power-law
continuum of photon index $\Gamma$, which is not included here. }
\end{figure}

XMM-Newton, which has a much larger effective area than ASCA, has now
observed MCG--6-30-15 and several other Seyfert 1 galaxies. The iron
line profile, kindly supplied by the PI of the observation, Joern
Wilms, is shown in Fig.~8. The data in this ratio plot resemble the
shape of the iron line from ASCA (Fig.~2). In the case of Mrk205
(Fig.~9) the line appears to peak around 6.7~keV, which may indicate
an ionized disc (i.e. the metals have only K-shell electrons). The
sharp peak at 6.4~keV may be a fluorescent line from more distant cold
gas within the galaxy. Models of ionized discs have been made for
constant density (e.g. Ross \& Fabian 1993) and more recently for
atmospheres in hydrostatic equilibrium (Nayakshin, Kazanas \& Kallman
2000; Ballantyne, Ross \& Fabian 2001: Fig.~7). An example of a
constant density model fitted to the ASCA MCG--6-30-15 data is shown
in Fig.~6. The models predict that the iron line shifts to 6.7~keV in
the rest frame and oxygen and other features should be present below
1~keV. Whether these low energy features are detectable depends on
their strength relative to the steep continuum.

\begin{figure}[] 
\hbox
{{\epsfig{file=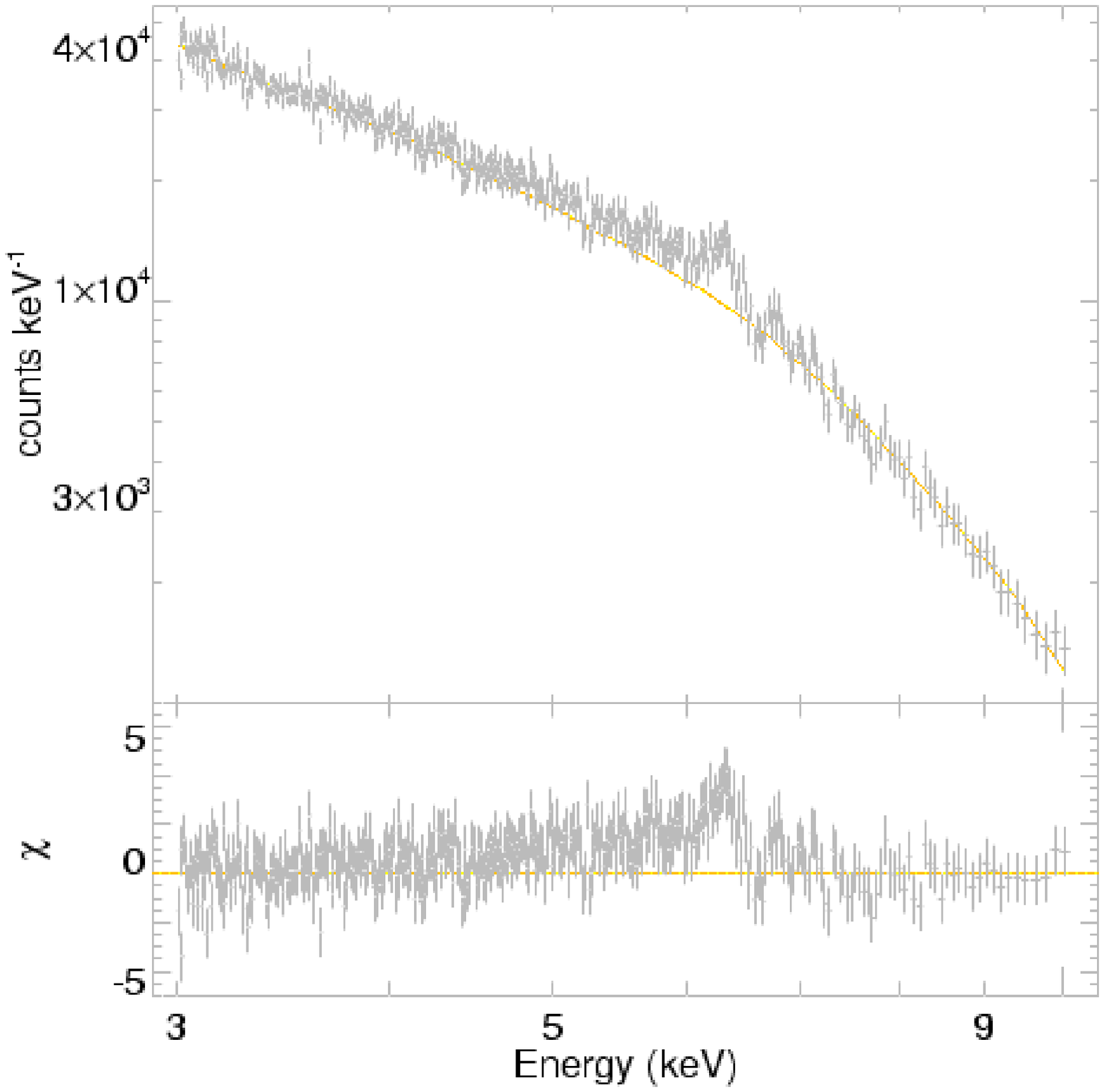,width=3in}}
{\epsfig{file=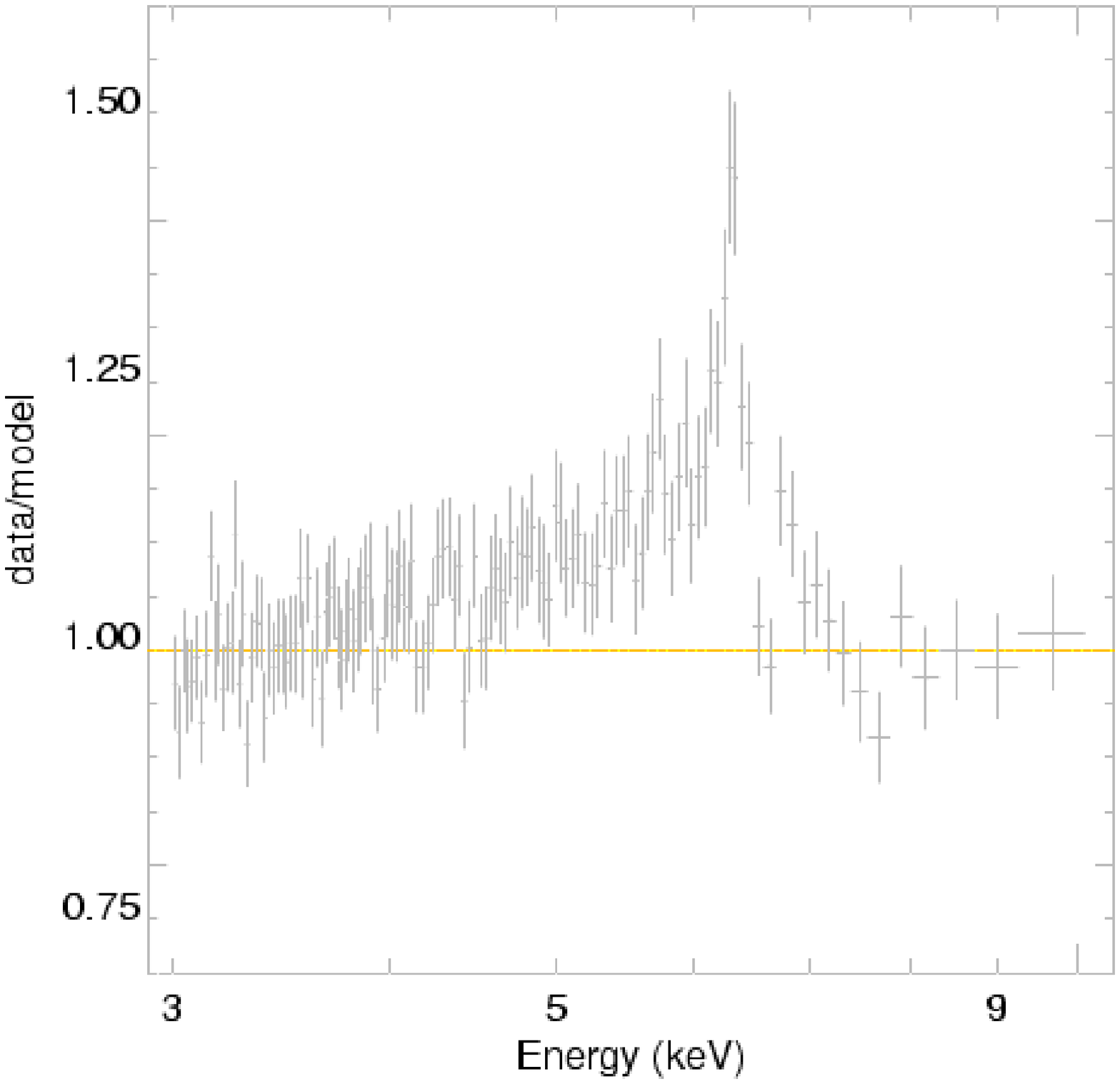,width=3in}}}
\vspace{10pt}
\caption{XMM-Newton EPIC-pn spectra of MCG--6-30-15. The left panel
shows the whole spectrum from 3--10~keV and the right panel shows the
ratio between the data and a simple power-law fitted only to the data
from 3--4 and 8--10~keV. Note that the ratio plot must be multiplied
by the power-law spectrum to produce a line intensity plot such as in
Fig. ~4. }
\end{figure}

\begin{figure}[] 
\hbox
{{\epsfig{file=mrk205_pn.ps,height=3in,width=3in,angle=270}}
{\epsfig{file=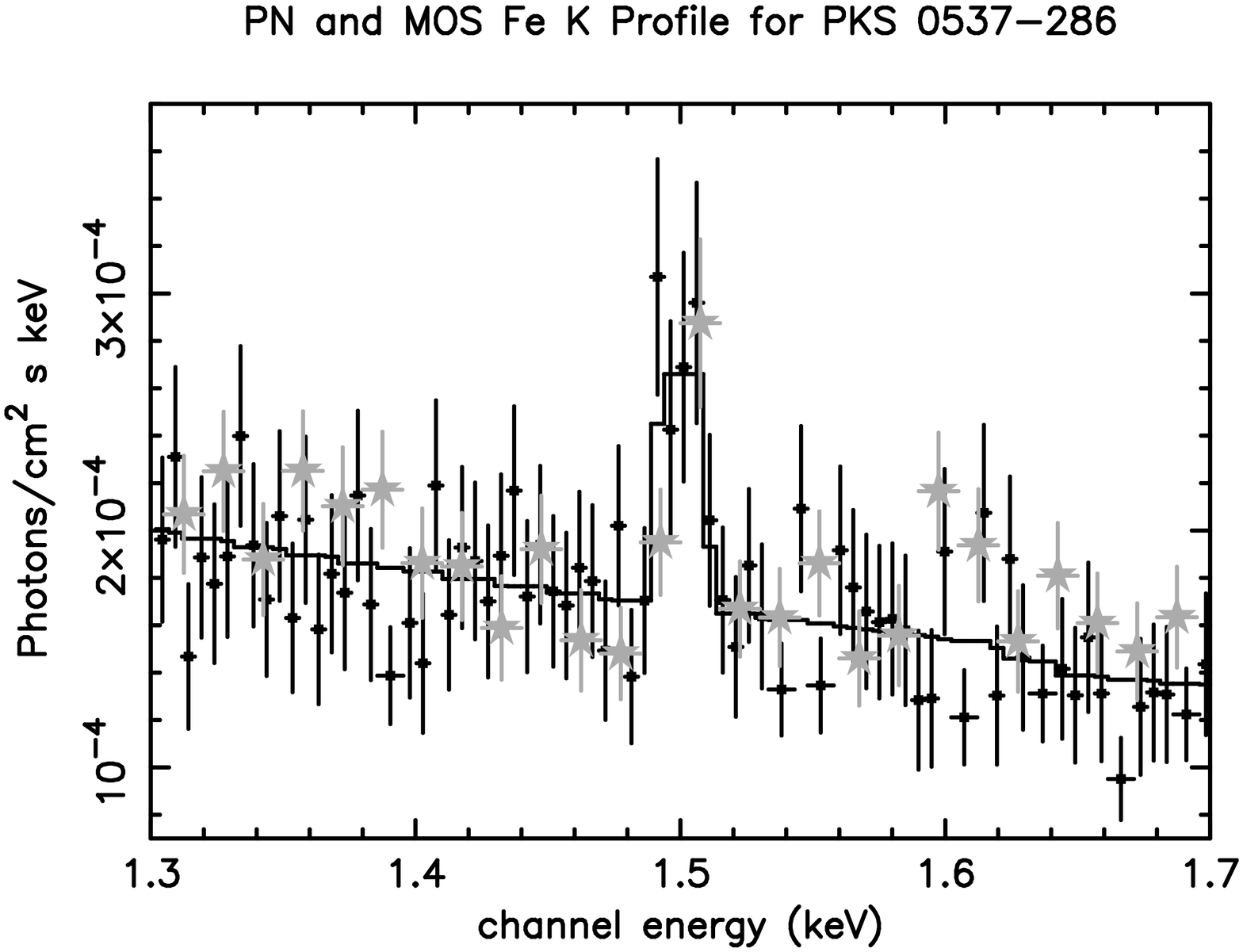,height=3in,width=3in,angle=270}}}
\vspace{10pt}
\caption{EPIC-pn spectra of the Seyfert 1 galaxy Mrk\,205 (left) and
the $z=3.104$ quasar PKS\,0537-286 (right), from Reeves et al (2001a,b).}
\end{figure}

Broad iron lines are generally not seen in objects with luminosity
exceeding a few time $10^{44}\ergps$ in the 2--10~keV band (Iwasawa \&
Taniguchi 1993; Nandra et al 1997). An example from XMM-Newton of a
narrow iron line from a powerful distant quasar, PKS\,0537-286, is
shown in Fig.~9, from Reeves et al 2001). Narrow iron lines are
expected in many AGN from reflection by outer gas at parsecs and
beyond (Ghisellini et al 1994) and are seen in some Seyferts (e.g.
Yaqoob et al 2000).

\begin{figure}[] 
\centerline{\epsfig{file=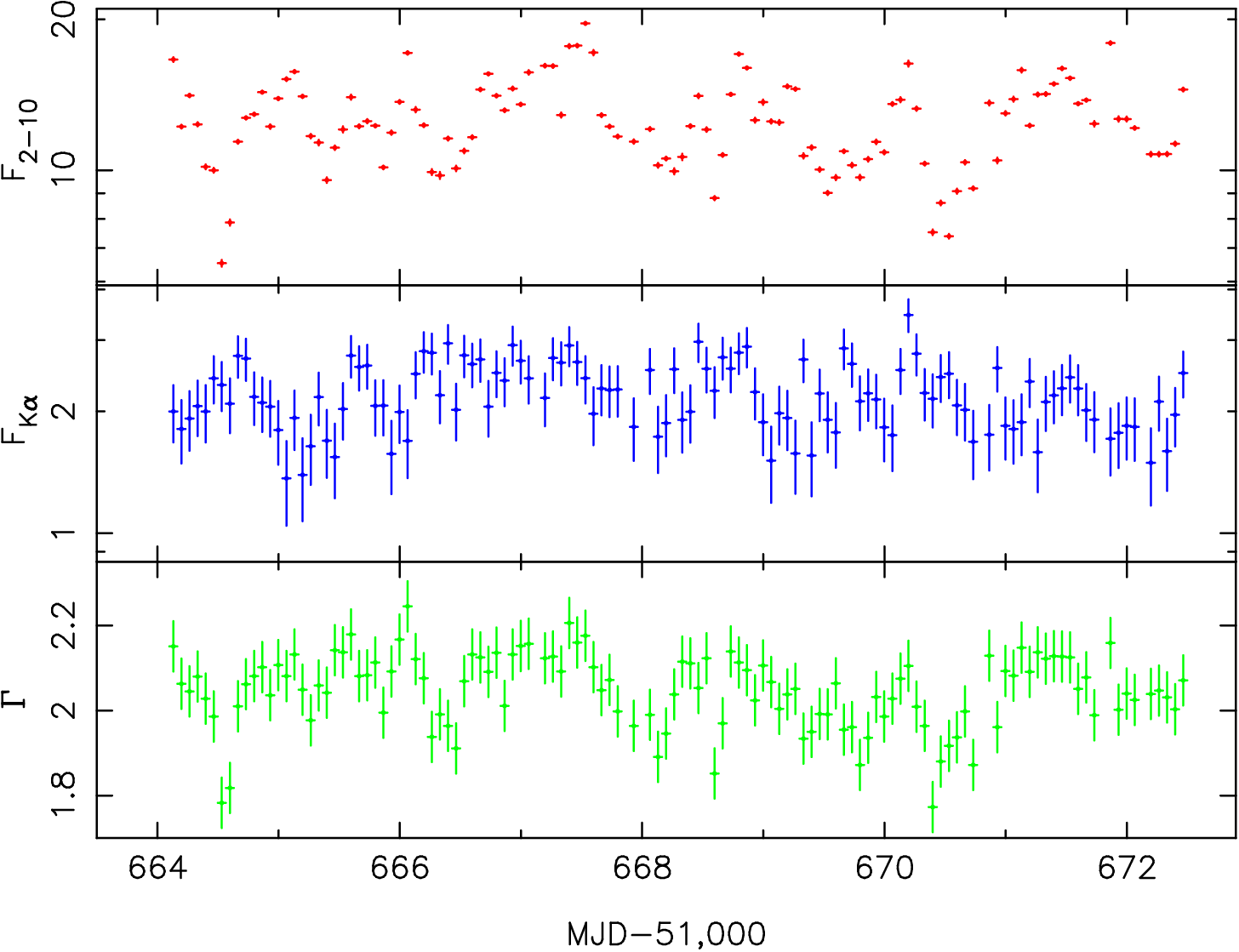,height=3.in,width=2.5in,angle=270}}
\centerline{\epsfig{file=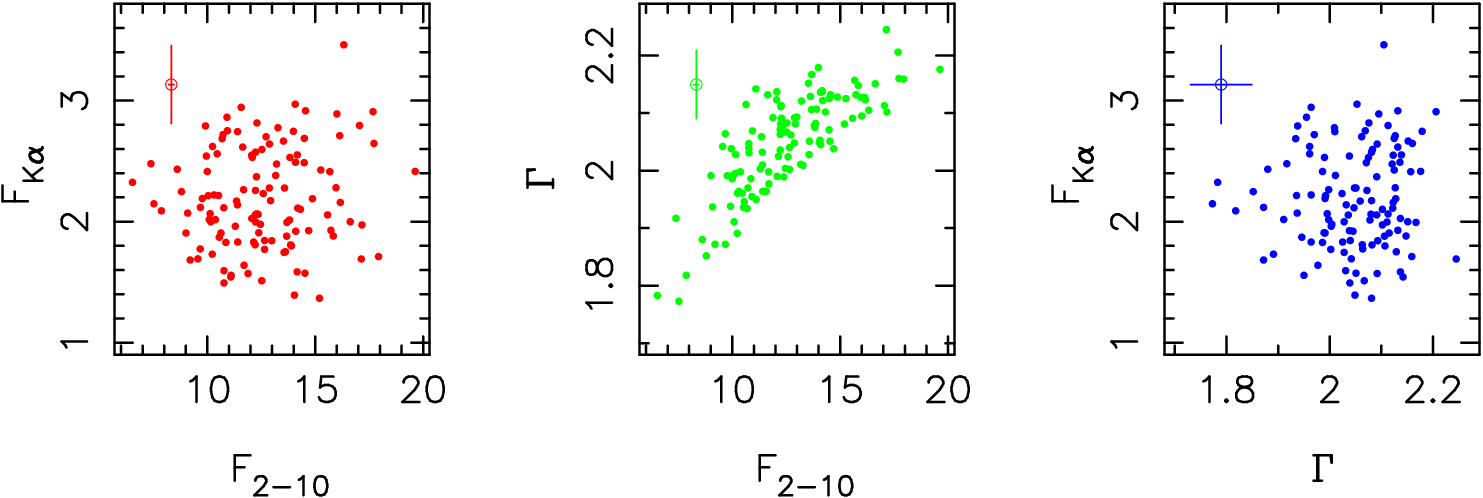,height=3.in,width=1.in,angle=270}}
\vspace{10pt}
\caption{Light curves of the total 2--10~keV band flux (top), iron
K-$\alpha$ flux (middle) and photon index (bottom) during the 1997
RXTE observation of MCG--6-30-15, from Vaughan \& Edelson (2000). The
lower panels show that there is no correlation between $F_{{\rm
K}\alpha}$ and $F_{2-10}$.}
\end{figure}

\begin{figure}[] 
\centerline{\epsfig{file=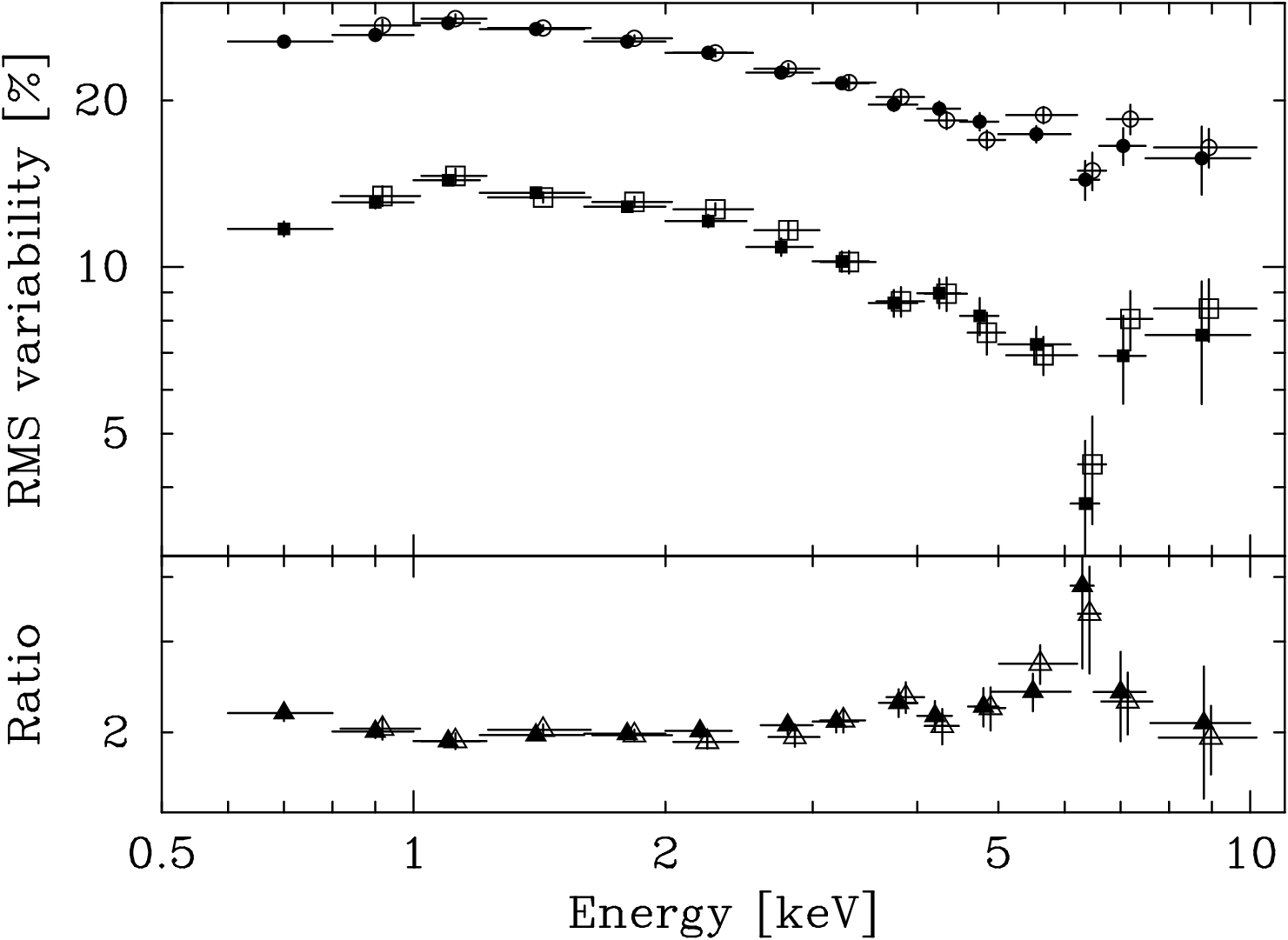,height=3in,width=2.5in,angle=270}}
\vspace{10pt}
\caption{The energy dependence of the RMS variability of MCG--6-30-15,
as observed with ASCA (from Matsumoto et al (2000). The 2 plots in the
top panel refer to time bins of (upper) $2\times 10^4\s$ and (lower)
$2\times 10^5\s$. }
\end{figure}

\section{Iron line variability}

The intensity of the iron line in MCG--6-30-15 does vary (Iwasawa et
al 1996; 99), but not in any obvious way in response to the continuum.
A long observation made simulataneously with ASCA and RXTE in 1997
(Lee et al 2000) confirms this. Vaughan \& Edelson (2000) have
re-analysed the RXTE data with the results shown in Fig.~10. There is
real variability in the iron line flux but it is not correlated with
the continuum flux. The continuum power-law slope is correlated in the
sense that the spectrum steepens as the flux increases. Further work
by Matsumoto et al (2000) on a 10 day ASCA observation in 1999 extends
the peculiar behaviour. The RMS variability decreases towards high
energy, particularly around the iron line (Fig.~11). 

Various ideas have been mooted in order to explain the variability.
Possibilities include; a) transrelativistic motion of flares (Reynolds
\& Fabian 1997; Beloborodov 1999) so that an observed bright continuum
is associated with flux beamed away from the disc and vice versa (this
has the problem that the spectral index trend with flux would be
opposite to that observed); b) smoke from the coronal flares
themselves (Merloni \& Fabian in prep.) in which electron scattering
in the corona smears out transmitted line emission when the regions
are spatially large, and hence bright; and c) ionization variations
(Lee et al 2000; Reynolds 2000).

The lack of any clear iron line -- continuum correlation so far
suggest that iron line reverberation (Reynolds et al 1999) may be
complicated.

\section{Broad oxygen lines?}

A recent development has been the observation of Seyfert galaxies at
high spectral resolution, provided by the gratings on Chandra and
XMM-Newton. Branduardi-Raymont et al (2001) claim from XMM-Newton RGS
spectra of MCG--6-30-15 the presence of relativistically-broadened
oxygen, nitrogen and carbon lines. As already noted, emission lines
from these elements are expected from an ionized disc. The surprise is
that they might be so strong. Moreover, the spectrum of the source
below 2 keV has been previously well modelled by a variable warm
absorber (Fabian et al 1994; Otani et al 1996). In their original
paper, Branduardi-Raymont et al (2001) rely on some unspecified
mechanism for the observed spectral turndown below 2 keV and argue
from the weak, broad ($2000\kmps$) absorption lines seen in their
spectra that the column density of any warm absorber is too low to
provide an edge large enough to account for the spectral jump at $\sim
0.7\keV$ (Fig. 12), moreover the jump occurs in the wrong place,
requiring a redshift of $16,000\kmps$.

Chandra HETG spectra provide a different view of the same object (Lee
et al 2001; Fig.~13). The 3 times higher spectral resolution shows that the
linewidths are $<200\kmps$ and, from curve of growth analysis of
the OVII line series, argues for the presence of a significant warm
absorber. Part of the redshift of the OVII photoelectric edge is then
accounted for by the convergence of the OVII absorption line series. 

The poster paper on the XMM-Newton RGS spectrum by Sako et al (2001;
these proceedings) now includes a warm absorber, together with
relativistically-broadened hydrogenic, C, N and O lines. Absorption by
neutral iron is suspected to be a further important ingredient by the
Chandra team. The neutral iron presumably being in the form of dust,
which must be present to explain the high optical/UV reddening seen in
the object (Reynolds et al 1997).

Since the conference, Lee et al (2001) have accounted for a major
feature around the OVII edge in terms of neutral iron absorption. Of
the three edges of iron-L, known as L3, L2 and L1, the absorption near
threshold of Fe L3 is a deep trough in laboratory data (Kortright \&
Kim 2000) and in the spectrum of Cyg X-1 (Schulz et al 2001). This
trough exactly matches the broad absorption feature seen in the
Chandra spectrum of MCG--6-30-15 at 0.71~keV (Fig.~13) and accounts
for the apparent redshift of the OVII edge (Fig.~12). Fe L2 and OVII
absorption together explain the remaining features between 0.7 and
0.75~keV.

Much more detailed modelling is required, covering a wider spectral
band, but it is now clear that MCG--6-30-15 does have a significant
dusty warm absorber. It is plausible that there is in addition some
emission from ionized C, N and O within the disc, but debatable as to
whether it is strong enough to create detectable features. The
relativistic line broadening parameters quoted by Branduardi-Raymont
et al (2001) do not yet agree with those found for the time-averaged
ASCA spectra. The ASCA results (Fig.~4) appear to be in accord with
the XMM-Newton EPIC data (Fig.~8) taken simultaneously with the RGS
spectra.

The Chandra spectrum does not show two spectral jumps, one of which can
be explained by photoelectric absorption and the other by the blue wing
of an emission line. I conclude that the large spectral jump seen
around 0.7~keV is dominated by absorption features from a dusty warm
absorber. Whether relativistically-broadened emission features are
also present remains to be seen.

\begin{figure}[] 
\centerline{\epsfig{file=mcg-6-30-15_data_en.ps,height=4.in,width=2.5in,angle=270}}
\vspace{10pt}
\caption{XMM-Newton RGS spectrum of MCG--6-30-15. The positions of the
OVIII and OVII photoelectric absorption edges are marked (see
Branduardi-Raymont et al (2001).}
\end{figure}


\begin{figure}[] 
\centerline{\epsfig{file=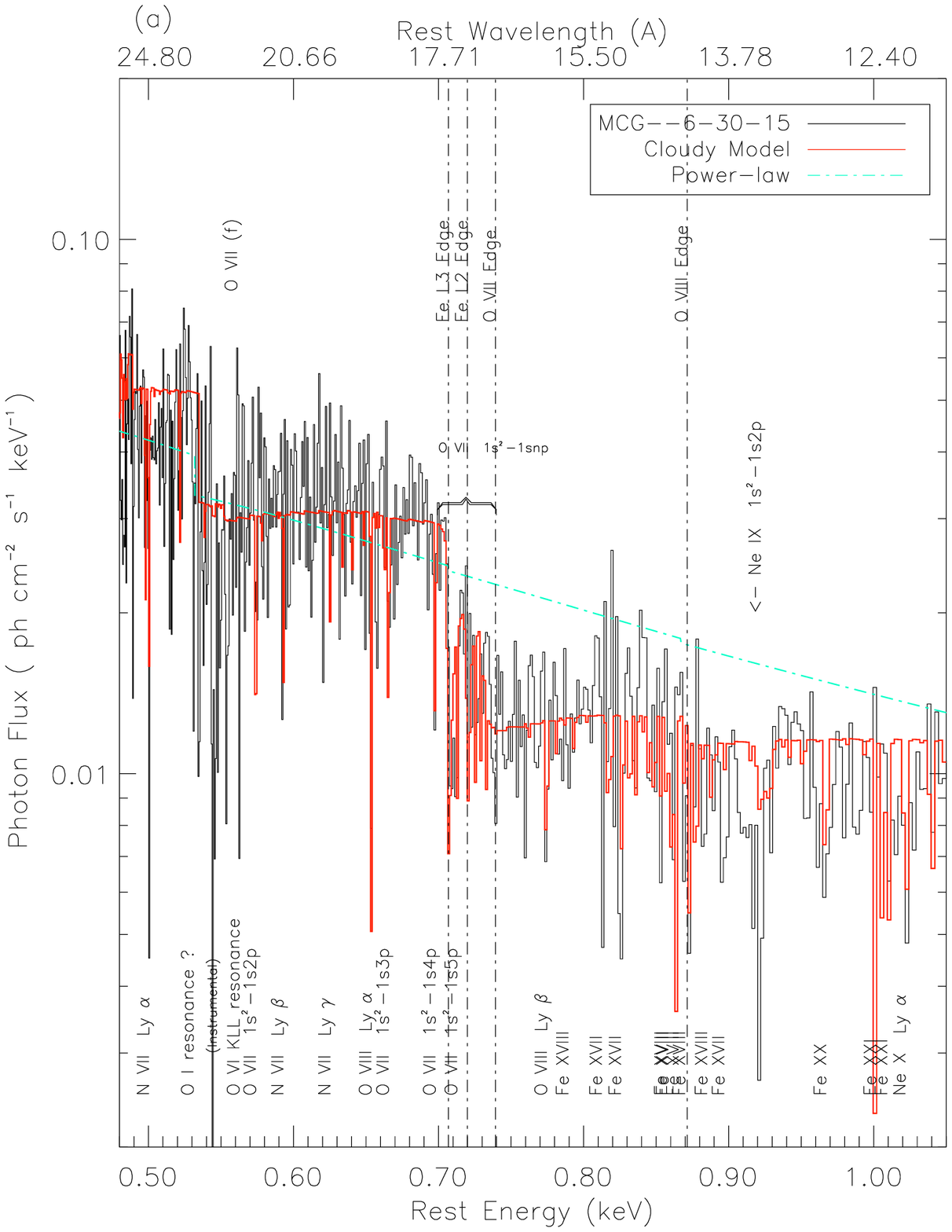,height=3.5in,width=7in}}
\vspace{10pt}
\caption{Chandra HETG spectrum of MCG--6-30-15 (Lee et al 2001). The CLOUDY model fit
includes absorption due to neutral Fe-L assumed to arise from dust in
the source.The large trough centred at 0.71~keV is due to Fe-L3
absorption.}
\end{figure}

\section{Summary}
The broad iron lines seen in MCG--6-30-15 and NGC\,3516 are well
explained by X-ray irradiation of an accretion disc extending very
close to a black hole. The skewed shape of the lines is mainly due to
the strong gravitational redshift in this region. 

The variability of the iron line and the lack of correlation with
continuum variations is not understood. The emission region is
plausibly complex with ionization, bulk motion and scattering effects
likely. 

The case for a broad oxygen line in MCG--6-30-15 is unclear.

\section{Acknowledgements}

I thank David Ballantyne, Claude Canizares, Kazushi Iwasawa, Julia
Lee, Raquel Morales, Patrick Ogle, Randy Ross, Norbert Schulz and
Simon Vaughan for help and Chiho Matsumoto, Paul Nandra, James Reeves,
Masao Sako and Joern Wilms for sending me plots in advance of
publication.

\end{document}